# Overview of Nonintercepting Beam-size Monitoring with Optical Diffraction Radiation

Alex H. Lumpkin

*Fermilab, Batavia IL 60510 USA*

**Abstract.** The initial demonstrations over the last several years of the use of optical diffraction radiation (ODR) as nonintercepting electron-beam-parameter monitors are reviewed. Developments in both far-field imaging and near-field imaging are addressed for ODR generated by a metal plane with a slit aperture, a single metal plane, and two-plane interferences. Polarization effects and sensitivities to beam size, divergence, and position will be discussed as well as a proposed path towards monitoring 10-micron beam sizes at 25 GeV.

## INTRODUCTION

The interest in nonintercepting beam-parameter monitors for high-average-power beams and/or high-areal-density beams in linear transport is growing as new facilities are being proposed with such properties. Early work on the theory of optical diffraction radiation (ODR) was reviewed in the late 1960's and early 1970's [1-2]. Application of ODR as a noninterceptive beam diagnostic began to be explored theoretically in the early 1980's [3-5], but the experimental demonstrations have been few, and only in the last decade. Developments in both far-field imaging and near-field imaging of optical diffraction radiation (ODR) are reviewed for ODR generated by a metal plane with a slit aperture, a single metal plane, and two-foil interferometers. Sensitivities to beam size, divergence, and offset position are somewhat convoluted, but techniques have been utilized at KEK [6], Frascati [7], and ANL [8] to separate in varying degrees such effects. KEK staff members have presented results at the 10- to 14-micron beam size regime using a slit configuration and pattern modulation visibility from an angle scan [6] and the Frascati group has used divergence, beam offset, and beam size in their fitting algorithms [7]. Focus-at-the-object or near-field ODR results for 7-GeV (APS), 4.5-GeV (JLAB), and 0.9-GeV (FLASH) electron beams have shown the feasibility of using this technique as a beam-size and position monitor [8-10]. In this case the polarized ODR image size and position *parallel* to the slit edge or single plane edge have shown reasonable sensitivities to these latter beam parameters. To date beam sizes from 1300 to 100 microns have been imaged. These results, when combined with preliminary modeling, support the proposed applications to future experiments such as the ILC test accelerator at Fermilab, CEBAF at JLAB, and FACET at SLAC. Polarization effects and sensitivities to beam size, divergence, and position will be presented as well as a proposed path towards monitoring 10-micron beam sizes at 25 GeV.

## EXPERIMENTAL AND ODR BACKGROUND

### Basic Imaging Aspects

Standard charged-particle beam imaging practices are used to characterize the beam properties in the accelerators at many labs over the years. The basic beam imaging system includes:

1) a conversion mechanism (scintillator, optical or x-ray synchrotron radiation (OSR or XSR) Cherenkov radiation (CR), undulator radiation (UR), optical transition radiation (OTR), and now optical diffraction radiation (ODR).

2) optical transport (port window, lenses, mirrors, band pass filters (BPF), polarizers)
3) imaging sensors such as a CCD or CID camera, with or without a microchannel plate image intensifier
4) video digitizer (for RS170 analog) or digital capture card (e.g., Firewire, camera link, Gig-E vision)
5) Image processing software.

A comprehensive summary on beam imaging aspects is provided by Bravin [11]. As a practical matter scintillators have good light conversion efficiency compared to OTR and ODR, but much longer response times. As particle properties evolved with the new photoelectric injector sources with the concomitant lower emittance beams, the need for a higher spatial resolution converter screen was identified. Since OTR is a surface phenomenon, it does not suffer from the finite volume and granularity aspects that result in limiting resolution terms in scintillators. In addition scintillators have a saturation aspect when one exceeds that threshold with higher charge areal densities. One can make trades on these aspects with OTR. These are all intercepting to the beam with the generation of beam scattering and bremsstrahlung, but OTR metal screen thicknesses of a few microns can be used instead of scintillator thicknesses of 100 μm. A complementary overview on OTR was provided at this workshop [12]. As mentioned in the INTRODUCTION however, we are looking for an option for a nonintercepting beam size monitor for high power beams with high charge areal density. One prime candidate is ODR.

## Basic ODR Issues

The intrinsic properties of ODR are similar to OTR. The radiation is emitted when a charged particle beam passes nearby a conducting screen, but does not intercept it. In this case the electromagnetic (EM) fields of the relativistic particle interact with the conducting screen to induce localized currents. These changing currents generate photons whose spatial distribution can be approximated by the method of virtual quanta as described in Jackson's classic book [13]. One generally considers the extent of the EM field as a flat circle of diameter $\gamma\lambda/2\pi$ (the ODR scaling parameter), where $\gamma$ is the Lorentz factor and $\lambda$ is the wavelength being detected by the sensor. The radiation intensity is approximately proportional to $\exp\{-2\pi a/\gamma\lambda\}$ where $a$ is the slit width. As could be expected, we have found empirically that the ODR source strength is strongly dependent on the impact parameter (IP), $d=a/2$, where $d$ is the distance from the beam to the screen edge. If $d$ is much larger than $\gamma\lambda/2\pi$, then no radiation is emitted: if approximately equal to it, one has measurable ODR, and if much less than it one approaches the OTR regime. In our experience, if we use an impact parameter of $d=5$ times the sigma-y of the beam, we have very little beam halo that strikes the foil. If one satisfies both relations, one has enough visible light to use a standard CCD camera on a single 3-nC micropulse as in reference [8]. In the case of the first experiments at $\gamma=2500$, however an angle scan was done with a PMT as sensor for a 1- nC micropulse [6].

Early treatments of ODR [3-5] generally considered the far-field imaging for an aperture in a metal plane which revealed a convolution of beam divergence, beam offset in the aperture, and beam size effects in the generated radiation from the metal. Separating these sensitivities is part of the challenge in making a beam-size monitor. The experiments to date have used a horizontal slit in a conducting metal plane to assess the vertical angular distribution pattern and its sensitivity to vertical beam size. Either one must configure the beam to have very low divergence such as at KEK or use the divergence as one of the fitting parameters as Frascati/FLASH collaborators have done [7]. There is an additional effect if the plane above and below the slit are displaced in angle or z and therefore slightly out of phase. These dephased planes may be used to provide sensitivity to beam size and alter the pattern [14, 15]. In addition, a second foil used as both a mask and source for interference effects has been tested now [15].

The experiments are summarized in Table 1. There are the two classes: 1) using far-field imaging and looking for angular distribution effects due to beam size and 2) using the alternative near-field or focus-at-the-object imaging of the ODR spatial distribution effects at the screen plane due to beam size. The main activities have been at KEK, Frascati/FLASH, and the author's collaborations on near-field ODR executed at APS, JLAB, and FLASH. The near-field model for the latter has been developed previously by D. Rule and reported [8].

**TABLE 1.** Summary of ODR experiments executed over the last several years using either far-field imaging of angular distributions or the alternative near-field imaging of spatial distributions techniques.

| Technique | Energy (GeV) | Beam Size (μm) | Charge (nC) | Detector | Div. (μrad) | Lab |
|---|---|---|---|---|---|---|
| Slit in plane, Far | 1.2 | 10-14 | 1 | PMT | 1.5 | ATF/KEK |
| Slit in plane, Far | 0.68 | 85 | 30 | Cooled CCD | 80 | INFN/FLASH |
| Single plane, Near | 7.0 | 1300 | 3 | CCD | 70 | APS/ANL |
| Single plane, Near | 4.5 | 120 | 3000 | CCD | 50 | FNAL/JLAB |
| Single plane, Near | 0.9 | 200 | 30 | Cooled CCD | 80 | FNAL/INFN |
| Two planes, Far | 0.9 | 90 | 30 | Cooled CCD | 80 | INFN/FLASH |

# EXPERIMENTAL RESULTS

One of the first experiments on beam size sensitivity in the far-field ODR was reported in 2004 [6]. The experimental work was initially performed at KEK ATF using a 260-µm x 5000 µm slit in a single, gold-coated Si screen. As seen in Fig. 1, the angular distribution pattern was obtained by scanning a mirror angle and directing either the OTR or ODR from the screen with a slit in it to a photomultiplier tube (PMT). This angle scan took about 10 minutes. The technique relied on the $I_{min}/I_{max}$ ratio in the angular distribution pattern to provide beam-size information. This size sensitivity depended on the very low 1.5 µrad divergence in the beam, which is a specialized case and generally not available. Larger divergences would have changed this ratio and complicated the analysis. As shown in Fig. 1b, the change in the ratio can be related to the change in $\sigma_y$ under these very low divergence conditions. The use of an upstream mask with a larger slit was needed to block upstream synchrotron light sources.

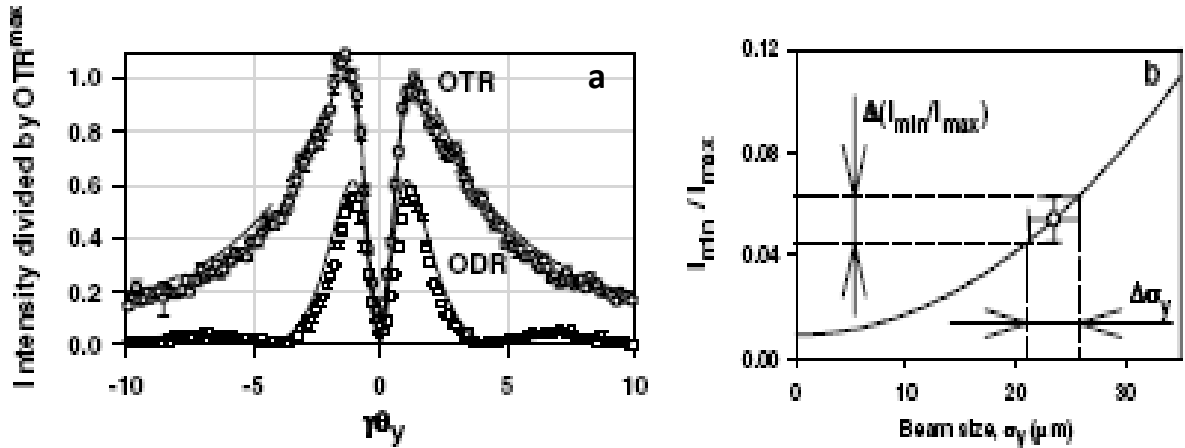

**FIGURE 1.** a) Comparison of the OTR and the ODR angular distribution patterns. In this case the peak intensity is 60% of the OTR. In b) the sensitivity of the $I_{min}/I_{max}$ ratio to vertical beam size is shown [6].

An additional series of experiments has been performed over the past few years using the accelerator at FLASH/DESY. These were motivated by evaluations performed by Castellano[4], an early experimental example of which was reported for 680 MeV beam by Chiadroni et al. [7], as noted in line 2 of Table 1.

The alternative technique is based on near-field imaging where one directly images the ODR spatial distribution at the screen surface as schematically shown in Fig. 2. This was a single metal plane of polished Al of 25 mm by 30 mm extent. The screen was mounted on an actuator linked to a stepper motor which allowed vertical positioning of the screen edge to an accuracy of 10 µm. The trajectory was confirmed with an rf BPM located just before the station and a downstream imaging screen. Initial tests included performing a scan of the edge screen position from 4 mm below the beam axis to 3mm above it. The downstream beam-loss monitor based on a Cherenkov detector was located just above the beam pipe and 1 m downstream. We established that when the edge was 4 to 5 sigma-y above the beam, the observed beam halo losses were minimal. This means that the predominant signal generated on the metal surface was ODR. The other background source is OSR from the upstream dipoles or quadrupoles. This was also evaluated to be only a broad distribution with a fraction of the intensity of the ODR [8]. Additionally, the OSR would be out of focus if generated 5 m upstream. In Fig. 3 we show the early comparison of the OTR and the ODR in the 7-GeV APS experiment. In this case the OTR was obtained with 8 times less micropulse charge at 0.4 nC. In subsequent tests, a remotely controlled filter wheel allowed us to select a neutral density filter to attenuate the OTR so we ran at the same beam charge conditions of 3 nC. The OTR-observed beam size was 1375 µm by 200 µm for x and y dimensions, respectively. The impact parameter was 6 sigma-y. This distance was still less than the nominal ODR scaling distance of 1.4 mm so sufficient strength ODR was generated. This matching of parameters allowed us to use a standard CCD camera for imaging the distribution on a single micropulse!

In the case of the JLAB experiments at 4.5 GeV, the beam sizes in the transfer line to the nuclear physics experiments were in the 125-300 µm regime [9]. The screen edge was inserted vertically under stepper motor control. The screen was an aluminized Si substrate prepared at FNAL. When the investigators used up to 80 µA CW beam with an impact parameter of 1 mm, the downstream loss monitors showed no detectable beam halo losses. The ODR signal levels were sufficient to explore wavelength effects and polarization effects as reported previously.

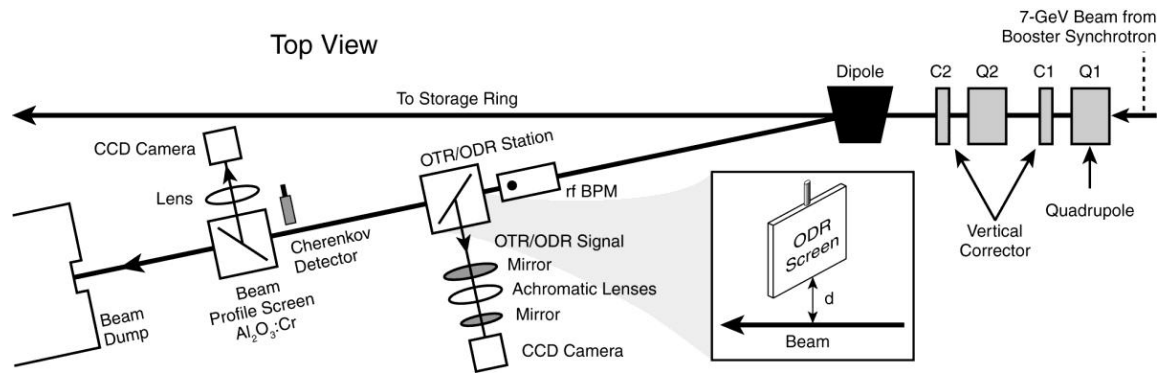

**FIGURE 2.** Schematic of the OTR/ODR experimental setup with near-field imaging at APS. (From Ref.8)

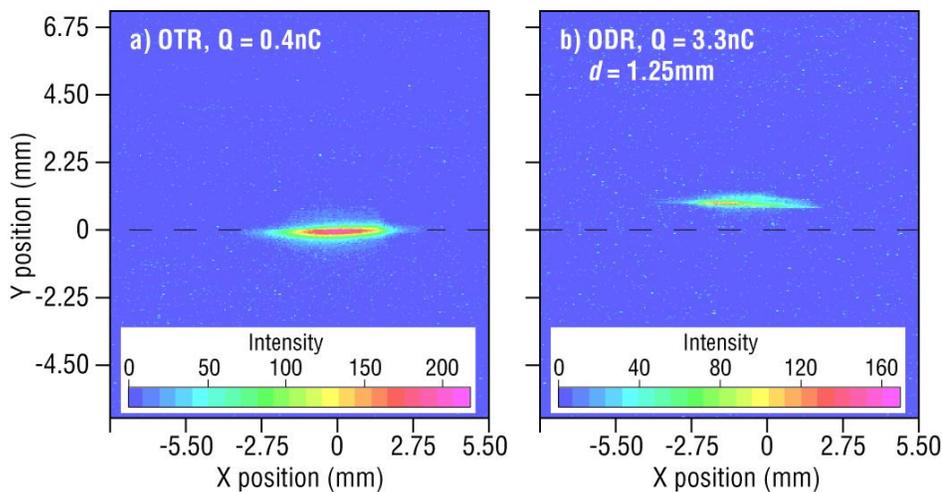

**FIGURE 3.** Early APS results showing a) the OTR image of the beam and b) the ODR image with impact parameter of 1.25 mm. The horizontal dashed line indicates the beam center position. Note the screen was inserted with a) its edge 4 mm below the centerline and b) 1.25 mm above the centerline. The ODR comes from the screen and its induced surface currents. (From Ref. 8)

The experiments at FLASH at 900 MeV were leveraged off the far-field angular distribution experiments mentioned earlier. In this case the lens was set to provide near field focus and an 800 x 40 nm BPF was used. The image was integrated over about 30 nC. With the beam slightly off center, the edges were 400 and 600 μm from the beam center. The actual horizontal size was 200 μm and the ODR image profile was about 375 μm, in qualitative agreement with our calculations [10]. The beam vertical dimension was 100 μm so the beam was still 4-6 sigma-y away in those units. However, because $\gamma\lambda/2\pi$ is 1.8 mm the induced currents were much lower, and the cooled CCD camera was the enabling detector that allowed imaging the ODR signal with only 30 nC integrated into an image.

## PROPOSED NEXT STEPS

An alternate way to display the status and the future directions that one might envision in the next 5 years for ODR investigations is shown in Fig. 4. In this case one tracks the nominal beam size addressed versus the beam energy. The demonstrated beam size studies listed in Table 1 are shown by black triangles. The proposed studies are shown with red circles. One future direction is indicated by the requirements expected in the International Linear Collider (ILC) [16]. Table 2 shows a comparison of some beam parameters at the ILC reference energies of 1, 5, 15, and 250 GeV. The SCRF test stand at NML or ILC-Test accelerator will provide ultimately the full average current of 9 mA at about 1 GeV. Calculations presented at BIW08 indicate potential ODR imaging at the 200-μm level for even a lower gamma=1000 beam [10]. Most of the ILC-like horizontal beam sizes are thus covered, but the smallest beam sizes need study. At this Workshop, the reported thrust of the laser wakefield accelerator (LWFA) towards the

10-GeV regime should not be ignored as another possible application for the diagnostic in the future. More immediately, in light of the description of the planned FACET facility [17], one can see that the parameter space is of great interest. Investigating the ODR beam-size sensitivity at 25 GeV has not been done to date, and this will directly address the challenge of small beam sizes at large gamma.

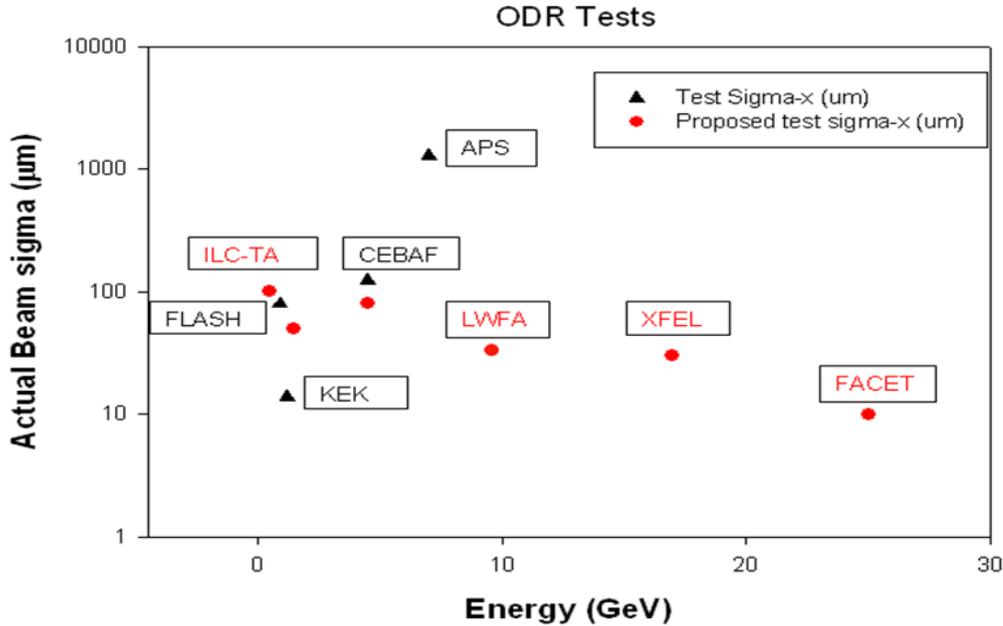

**FIGURE 4.** Summary chart of demonstrated ODR test beam sizes versus electron beam energy. The previously demonstrated tests (black triangles) and the proposed tests (red circles) are indicated. The FACET parameters offer a new regime to explore.

Experiments at FACET have been proposed to use the near-field technique to investigate the feasibility of attaining beam size sensitivity at the 10-μm regime and for a beam at 25 GeV. Initial calculations shown in Fig. 5 address this question for an impact parameter of 50 μm. For the perpendicular polarization component the ODR profile changes noticeably for 20 and 35 μm input sizes in the model, but this measurement is challenging at 10 μm as shown in Fig. 5a. Alternatively, according to our model, the parallel polarization component actually has a stronger sensitivity in the $I_{min}/I_{max}$ ratio for the change from 10 to 20 μm as seen in Fig. 5b. The valley minimum changes from 20% to 50% of the peak intensity, and this should be measurable. However, this component is 3-4 times weaker so we may need to implement a more sensitive camera option to enhance video signal levels and statistical aspects.

**TABLE 2.** Summary of parameters of interest for ILC [16] and comparison to those expected at NML/Fermilab and FACET/SLAC. Note the high average current proposed for ILC-TA and the ILC.

| Parameter | ILC-TA/NML | FACET | ILC |
|---|---|---|---|
| Energy (GeV) | 1 | 25 | 1, 5, 15, 250 |
| X beam size (μm) | 200 | 10 | 650, 300, 150, 30 |
| Y beam size (μm) | 80 | 10 | 35, 15, 8, 2 |
| Current (mA) | 9 | 3 x 10e-6 | 9 |

# SUMMARY

In summary, I have reviewed the experimental progress on beam size monitoring over the last several years using nonintercepting ODR techniques. Successful results have been obtained with initially only far-field angular distribution measurements, but now the near-field results have been obtained at three labs and three different energies. These include the 80 µA CW tests at CEBAF/JLAB with a 4.5-GeV energy and 120-µm beam sizes. It has also been proposed that the ILC-TA/NML/FNAL and FACET/SLAC parameters will be ideal next steps for exploring ODR applicability to high average power and high beam energy, respectively. In the next five years further tests with the polarization components, wavelength effects, masking, and point-spread-function deconvolution procedures [18] would also be useful in mapping the parameter space of the technique's applicability.

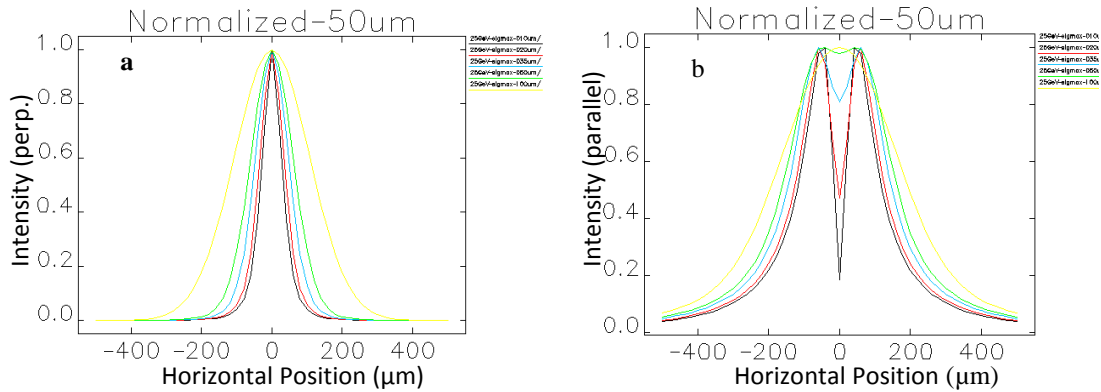

FIGURE 5. Numerical simulations of the near-field ODR projected horizontal profile sizes for various input x beam sizes: The 10-, 20-, 35-, 50-, and 100-µm input cases are the black, red, blue, green, and yellow curves, respectively, with IP=50µm for FACET: a) perpendicular polarized component (perp) and b) parallel polarization component (parallel). The vertical beam size was held at 10 µm, and the wavelength used in the simulation was 0.8 µm.

## ACKNOWLEDGMENTS

The author acknowledges his ODR collaborations with C.-Y. Yao, N. Sereno, S. Pasky, Y. Li, and W. Berg at ANL; D.W. Rule of NSWC; P. Evtushenko, A. Freyberger, and C. Liu at JLAB; and E. Chiadroni and M. Castellano of INFN/Frascati; A. Cianchi of Univ. of Rome Tor Vergato; and K. Hankavarra of FLASH/DESY. He also acknowledges support from M. Wendt and M. Church of FNAL. This work was supported by the Fermi Research Alliance, LLC under Contract No. DE-AC02-07CH11359 with the U.S. Department of Energy.